\documentclass[english,sort,largeformat,twocolumn,final]{interact}
\usepackage{epstopdf}

\usepackage[numbers,sort&compress,merge]{natbib}

\usepackage[T1]{fontenc}
\usepackage[utf8]{inputenc}
\usepackage{xcolor}
\usepackage{pdfcolmk}
\usepackage{refstyle}
\usepackage{rotfloat}
\usepackage{booktabs}
\usepackage{amsmath}
\usepackage{mathabx}
\usepackage{graphicx}
\PassOptionsToPackage{normalem}{ulem}
\usepackage{ulem}
\usepackage{url}

\makeatletter

\AtBeginDocument{\providecommand\eqref[1]{\ref{eq:#1}}}
\AtBeginDocument{\providecommand\tabref[1]{\ref{tab:#1}}}
\AtBeginDocument{\providecommand\figref[1]{\ref{fig:#1}}}
\providecommand{\tabularnewline}{\\}
\RS@ifundefined{subsecref}
  {\newref{subsec}{name = \RSsectxt}}
  {}
\RS@ifundefined{thmref}
  {\def\RSthmtxt{theorem~}\newref{thm}{name = \RSthmtxt}}
  {}
\RS@ifundefined{lemref}
  {\def\RSlemtxt{lemma~}\newref{lem}{name = \RSlemtxt}}
  {}

\providecolor{lyxadded}{rgb}{0,0,1}
\providecolor{lyxdeleted}{rgb}{1,0,0}

\usepackage{chemfig}
\usepackage[version=3]{mhchem}
\usepackage{braket}
\usepackage{placeins}

\@ifundefined{showcaptionsetup}{}{%
 \PassOptionsToPackage{caption=false}{subfig}}
\usepackage{subfig}
\makeatother

\usepackage{babel}
\begin{document}
\title{Orbital optimization in the perfect pairing hierarchy. Applications
to full-valence calculations on linear polyacenes}
\author{
  \name{Susi Lehtola\textsuperscript{a}\thanks{S.L. Email: susi.lehtola@alumni.helsinki.fi} and John Parkhill\textsuperscript{b}\thanks{J.P. Email: john.parkhill@gmail.com} and Martin Head-Gordon\textsuperscript{a,c}\thanks{M.H.G. Email: mhg@cchem.berkeley.edu}}
  \affil{\textsuperscript{a}Chemical Sciences Division, Lawrence Berkeley National Laboratory, Berkeley, California 94720, United States; \textsuperscript{b}Department of Chemistry, University of Notre Dame, 251 Nieuwland Science Hall, Notre Dame, Indiana 46556, United States; \textsuperscript{c}Department of Chemistry, University of California, Berkeley, California 94720, United States}
  }

\maketitle

\begin{abstract}
We describe the implementation of orbital optimization for the models
in the perfect pairing hierarchy [Lehtola et al, J. Chem. Phys. 145,
  134110 (2016)]. Orbital optimization, which is generally necessary
to obtain reliable results, is pursued at perfect pairing (PP) and
perfect quadruples (PQ) levels of theory for applications on linear
polyacenes, which are believed to exhibit strong correlation in the
$\pi$ space. While local minima and $\sigma$-$\pi$ symmetry breaking
solutions were found for PP orbitals, no such problems were
encountered for PQ orbitals. The PQ orbitals are used for single-point
calculations at PP, PQ and perfect hextuples (PH) levels of theory,
both only in the $\pi$ subspace, as well as in the full $\sigma\pi$
valence space. It is numerically demonstrated that the inclusion of
single excitations is necessary also when optimized orbitals are
used. PH is found to yield good agreement with previously published
density matrix renormalization group (DMRG) data in the $\pi$ space,
capturing over 95\% of the correlation energy. Full-valence
calculations made possible by our novel, efficient code reveal that
strong correlations are weaker when larger bases or active spaces are
employed than in previous calculations. The largest full-valence PH
calculations presented correspond to a (192e,192o) problem.
\end{abstract}


\abbreviations{AO: atomic orbital, BFGS:
  Broyden–Fletcher–Goldfarb–Shanno, CAS-SCF: complete active space
  self-consistent field, CCVB: coupled cluster valence bond, CI:
  configuration interaction, DMRG: density matrix renormalization
  group, FCI: full configuration interaction, GDM: geometric direct
  minimization, HONO: highest occupied natural orbital, LUNO: lowest
  unoccupied natural orbital, MC-SCF: multiconfigurational
  self-consistent field, MO: molecular orbital, NO: natural orbital,
  NOON: natural orbital occupation number, OO-CC: orbital-optimized
  coupled cluster, OPDM: one particle density matrix, PP: perfect
  pairing, PQ: perfect quadruples, PH: perfect hextuples, PPH: perfect
  pairing hierarchy, TPDM: two particle density matrix, VDRM:
  variational reduced density matrix, VOO-CC: valence
  orbital-optimized coupled cluster}

\global\long\def\ERI#1#2{(#1|#2)}
\foreignlanguage{english}{}\global\long\def\bra#1{\Bra{#1}}
\foreignlanguage{english}{}\global\long\def\ket#1{\Ket{#1}}
\foreignlanguage{english}{}\global\long\def\braket#1{\Braket{#1}}

\selectlanguage{english}%
\newcommand*\citeref[1]{ref. \citenum{#1}}
\newcommand*\Citeref[1]{Ref. \citenum{#1}}

\section{Introduction}

Systems with strong correlation, in which many electronic
configurations contribute significantly to the wave function, are
important in many areas of chemistry such as catalysis. Unfortunately,
their accurate yet efficient modeling is still an unsolved question in
theoretical chemistry. The exact wave function for any system is
available in theory by diagonalizing the molecular Hamiltonian in the
basis of electron configurations, yielding the full configuration
interaction (FCI) approach. But, FCI exhibits exponential scaling that
limits its use to tiny systems \cite{Knowles1989, Knowles1989a,
  Rossi1999, Thogersen2004, Gan2005, Gan2006}.

Restricting the number of electronic configurations allowed in the
wave function yields the multiconfigurational self consistent field
(MC-SCF) method \cite{Hinze1973}. However, as typically a considerable
number -- thousands to millions -- of configurations are necessary to
obtain a sufficiently converged wave function, the electronic
configurations allowed in the wave function are more often defined in
terms of active orbitals, as in the complete active space
self-consistent field (CAS-SCF) method \cite{Siegbahn1980, Roos1980,
  Siegbahn1981} or variants thereof \cite{Olsen1988, Fleig2001,
  Ma2011}. CAS-SCF still maintains the exponential scaling of FCI with
respect to the size of the active space, and is generally acknowledged
to be limited to problems around the size of 16 electrons in 16
orbitals, denoted as (16e,16o). While ideally all valence electrons
should be included in the active space, due to the limitation on the
size of the feasible active space the choice of the active orbitals is
generally a complicated problem, and improper choices often yield
unreliable results \cite{Veryazov2011}. The necessary active space can
also change e.g. along a reaction path, further underlining the
problems caused by the limitations of this type of an approach.

Because most of the configurations in an FCI (or CAS-SCF) wave
function typically have negligible weights, it is possible to truncate
the wave function without losing significant accuracy
\cite{Ivanic2001, Ivanic2002, Bytautas2003, Bytautas2009}. This has
been recently been exploited in various stochastic and adaptive
approaches to FCI and CAS-SCF \cite{Booth2009, Booth2010, Booth2011,
  Shepherd2012, Shepherd2012a, Shepherd2012b, Daday2012, Thomas2015,
  Schriber2016, Tubman2016, Zhang2016a, Holmes2016, Sharma2017}, which
have made quasi-FCI calculations feasible on much larger systems than
before. For instance, \citeref{Sharma2017} describes the solution of
the (28e,22o) problem in oxoMn-salen to sub-millihartree accuracy in
less than a minute on 20 cores. Although these novel approaches for
the solution of the FCI problem have made calculations possible on
much larger systems than with conventional algorithms, due to their
reliance on the FCI ansatz, they still scale exponentially in the size
of the active space.

The density matrix renormalization group (DMRG) approach
\cite{White1992, Chan2011, Marti2011, Wouters2014b,
  Olivares-Amaya2015} is able to reproduce the FCI result, yet it
affords polynomial scaling in linear systems. Unfortunately, for
non-linear systems, like CAS-SCF or FCI, DMRG succumbs to exponential
scaling as well \cite{Olivares-Amaya2015}. Still, DMRG has successfully
been applied to considerably larger active spaces than CAS-SCF
\cite{Kurashige2009,Kurashige2013,Wouters2014a}, and is generally
thought to be the best general-purpose tool for strong correlation as
it can be used to treat extremely challenging problems such as
open-shell transition metal compounds. In practice, DMRG is limited to
($n_{\text{el}}$e,$n_{\text{orb}}$o) problems with $n_{\text{el}}
n_{\text{orb}} \leq 2000$ \cite{Olivares-Amaya2015}. Furthermore, DMRG
is not trivial to use due to issues with the choice of various
convergence parameters. There is thus still demand for approximate
methods that are able to capture the bulk of strong correlation
effects while also being fast enough to be able to handle a
full-valence active space. The perfect pairing hierarchy (PPH) is one
possible candidate for such a method.

The methods in the PPH are obtained by truncating the equations of
infinite-order coupled cluster (CC) theory \cite{Cizek1966} to be
exact for a singlet state (which can be open- or closed-shell) in a
($n$e,$n$o) active space for a given number of electrons $n$
\cite{Parkhill2009, Parkhill2010}. The truncation of PPH bears some
similarity to, but is not to be confused with the $n$CC hierarchy
proposed by Bartlett and Musia\l{} \cite{Bartlett2006}, where the CC
\emph{diagrams} are truncated for exactness for $n$ electrons. Thus,
the PPH can alternatively be understood as a further truncation of
$n$CC to an active space.

The simplest model in the PPH, perfect pairing \cite{Hurley1953,
  Hunt1972, Ukrainskii1977, Goddard1978, Cullen1996, Beran2005} (PP),
is exact for a $n=2$ singlet, whereas the perfect quadruples
\cite{Parkhill2009} (PQ) and perfect hextuples \cite{Parkhill2010}
(PH) models are exact for $n=4$ and $n=6$, respectively. The PP, PQ,
and PH models form a hierarchy of approximations to CAS-SCF. With our
recent, efficient implementation of the PQ and PH models
\cite{Lehtola2016b}, this hierarchy can be used to perform approximate
CAS-SCF calculations on systems of unforeseen size, as will be seen
later on in the manuscript.

While CC theory is invariant to orbital rotations in the
occupied-occupied and virtual-virtual blocks, the results of truncated
CC approaches generally depend on the set of occupied orbitals used in
the calculation. In orbital-optimized CC (OO-CC) theory
\cite{Sherrill1998} occupied-virtual (ov) rotations are performed to
minimize the CC energy. Furthermore, as core orbitals are chemically
inactive, an active space can be introduced, as is also done in
CAS-SCF, yielding the valence orbital-optimized CC (VOO-CC) method
\cite{Krylov1998}. The introduction of an active space necessitates
optimization of rotations between active and inactive occupied
orbitals, as well as active and inactive virtual orbitals, in addition
to the ov rotations in OO-CC. In the PPH, mixings \emph{within} the
active occupied, as well as active virtual orbital spaces must further
be considered, because the truncation is based on the orbitals.

In the present manuscript, we describe how orbital optimization can be
efficiently implemented for the models within the PPH, and present
their applications to full-valence calculations on linear polyacenes.
Ever since preliminary calculations predicted linear polyacenes to
have singlet diradical ground states \cite{Bendikov2004}, both linear
and cyclacenes have been studied intensively using a variety of
methods \cite{Bettinger2010}. Famously, DMRG calculations have been
used to show that the larger linear polyacenes exhibit strong
correlation for their $\pi$ electrons \cite{Hachmann2007}. Other
approaches used have included density functional theory
\cite{Jiang2008a, Ess2011, Chai2012, Torres2014, Bettinger2016,
  Rayne2016}, multiconfiguration pair-density functional theory
\cite{Ghosh2017}, projected Hartree–Fock theory \cite{Rivero2013}, the
random phase approximation \cite{Yang2016}, configuration interaction
\cite{Bettinger2016} (CI), adaptive CI \cite{Schriber2016}, GW theory
\cite{Wilhelm2016}, variational two-electron reduced density matrix
(VRDM) theory \cite{Gidofalvi2008, Pelzer2011, Fosso-Tande2015,
  Fosso-Tande2016}, Møller–Plesset perturbation theory
\cite{Deleuze2003, Hajgato2009, Hajgato2011, Yost2016, Rayne2016},
spin-flip methods \cite{Casanova2009a, Ibeji2015}, CAS-SCF
\cite{Bendikov2004, Hajgato2009, Hajgato2011, Torres2014,
  Battaglia2017} as well as restricted-active space self-consistent
field theory \cite{Aiga2012}, valence bond \cite{Qu2009} and CC
valence bond (CCVB) theory \cite{Small2014, Lee2017}, CC theory
\cite{Deleuze2003, Hajgato2009, Hajgato2011, Lee2017} and
multireference averaged quadratic CC theory \cite{Plasser2013,
  Horn2014}, an algebraic diagrammatic construction scheme
\cite{Knippenberg2010}, as well as PPH methods with approximate
orbitals \cite{Lehtola2016b}.

We chose these systems for our study, because the correlation
exhibited is too strong for successful description with conventional
CC approaches, such as CC with full single and double excitations
(CCSD) \cite{Hajgato2009, Deleuze2003, Lee2017}, and because the
CAS-SCF, VRDM, and DMRG studies have been restricted to the $\pi$
space only, as full-valence calculations would otherwise be too costly
for these methods for anything beyond the smallest acenes.  As far as
we are aware, the only full-valence calculations on large polyacenes
employing a method potentially capable of accurate treatment of strong
correlation that have been published are the CCVB calculations
performed contemporaneously in our group \cite{Small2014,
  Lee2017}. Thanks to our recent implementation \cite{Lehtola2016b},
even the larger polyacenes can be treated accurately in full-valence
active spaces using PPH methods, yielding an independent way to check
the correctness of the recent CCVB results.

As the result of our application, we find that orbital optimization
can be routinely applied to large systems, and that it significantly
improves the accuracy of the PPH models. We show that even when
optimal orbitals are used, it is necessary to include single
excitations as the optimal orbitals are not Brueckner orbitals. We
show that PH successfully describes strong correlation in the STO-3G
$\pi$ space compared to literature DMRG reference values
\cite{Hachmann2007}. In agreement with the recent CCVB results of
\citeref{Lee2017}, and in contrast to the $\pi$ active space results
obtained using e.g. DMRG \cite{Hachmann2007} and VDRM
\cite{Gidofalvi2008, Fosso-Tande2015, Fosso-Tande2016}, we show the
strong correlation is quenched when all valence orbitals are included
in the active space. Furthermore, going beyond the STO-3G basis for
the first time with a high-level full-valence method shows that
improvement of the basis set leads to even more quenching of the
strong correlation for the same active space.


In the following, $i$, $j$ and $k$ denote
occupied spin-orbitals, $a$ and $b$ denote unoccupied spin-orbitals,
and $p$, $q$, $r$, $s$, $t$ and $u$ denote general spin-orbitals.

\section{Theory}

Analytic gradients and orbital-optimized CC theory have been discussed
extensively elsewhere \cite{Sherrill1998, Krylov1998, Bozkaya2011},
but a brief overview of the optimization procedure is given below for
the sake of completeness. The CC equations are given by
\begin{align}
\braket{\Phi_{0}|\overline{H}|\Phi_{0}}= & E_{c},\label{eq:CC-E}\\
\braket{\Phi_{i}^{a}|\overline{H}|\Phi_{0}}= & 0,\label{eq:CC-S}\\
\braket{\Phi_{ij}^{ab}|\overline{H}|\Phi_{0}}= & 0,\label{eq:CC-D}\\
\vdots\nonumber
\end{align}
where $\overline{H}$ is the effective Hamiltonian, the correlation
energy $E_{c}$ is given by \eqref{CC-E}, and \eqref{CC-S,CC-D}
determine the single and double excitation ampitudes,
respectively. Regardless of the truncation of the amplitudes, the CC
energy is determined solely by the single and double excitations as
\begin{align}
E^{\text{CC}}= & E_{0}+\sum_{ia}f_{i}^{a}t_{i}^{a}+\frac{1}{2}\sum_{ijab}t_{i}^{a}t_{j}^{b}\braket{ij||ab}+\frac{1}{4}\sum_{ijab}t_{ij}^{ab}\braket{ij||ab},\label{eq:E-CC}
\end{align}
where $E_{0}$ is the Hartree–Fock reference energy, $f_{i}^{a}$ is a
Fock matrix element and $\braket{ij||ab}$ is a two-electron integral, and $t_{i}^{a}$ and $t_{ij}^{ab}$ are the single
and double excitation amplitudes, respectively.

The orbitals enter the problem through the matrix elements $f_{i}^{a}$
and $\braket{ij||ab}$.  However, the orbital rotation gradient cannot
be calculated from \eqref{E-CC}, because the amplitudes $t$ depend on
the matrix elements in the first order. To be able to optimize the
orbitals, one must demand the simultaneous satisfaction of the CC
equations through Lagrangian multipliers $\lambda$ \cite{Koch1990a}
\begin{align}
\mathcal{L}= & E^{\text{CC}}+\sum_{ia}\lambda_{i}^{a}\braket{\Phi_{i}^{a}|\overline{H}|\Phi_{0}}+\sum_{ijab}\lambda_{ij}^{ab}\braket{\Phi_{ij}^{ab}|\overline{H}|\Phi_{0}}+\dots\label{eq:L}
\end{align}
that are known as de-excitation amplitudes, in analogy to the
excitation amplitudes $t$.

Demanding that the variation of the CC Lagrangian (\eqref{L}) vanish
both with respect to the $t$ and the $\lambda$ amplitudes, equations
are obtained for the $\lambda$ and the $t$ amplitudes,
respectively. Having solved for the values of $t$ and $\lambda$, the
terms in \eqref{L} can be regrouped into the standard form
\begin{align}
\mathcal{L}= & E_{0}+\gamma_{pq}f_{pq}+\Gamma_{pqrs}\braket{pq||rs},\label{eq:E-DM}
\end{align}
where $\gamma_{pq}$ and $\Gamma_{pqrs}$ are then identified as the
one-particle (OPDM) and two-particle density matrices (TPDM),
respectively. Unlike the CC energy (\eqref{E-CC}), the CC Lagrangian
(\eqref{L, E-DM}) is correct to the first order in the matrix
elements. Thus, $\partial\mathcal{L}/\partial\theta_{pq}$ is a proper
gradient for a rotation of the orbitals through an angle
$\boldsymbol{\theta}$
\begin{align}
\boldsymbol{C}\to & \boldsymbol{C}e^{\boldsymbol{\theta}}. \label{eq:orbital-rotation}
\end{align}
The in-detail equations for the gradient are given in the Appendix.

\section{Computational Details}

The geometric direct minimization (GDM) approach
\cite{VanVoorhis2002}, which uses a limited-memory
Broyden–Fletcher–Goldfarb–Shanno (BFGS) approximate Hessian
\cite{Nocedal1999}, is used to optimize the orbitals
\cite{VanVoorhis2002a, Beran2005}. No parallel transport of previous
gradients in the BFGS is used in our implementation, as we are viewing
the optimization problem on the Stiefel manifold where the efficient
implementation of parallel transport is an unsolved problem
\cite{Edelman1998}. To speed up the convergence of the optimization,
also the diagonal Hessian $\partial^{2} \mathcal{L} / \partial
\theta_{pq}^{2}$ is calculated (see the Appendix) and used to
initialize the BFGS solver.

Once a search direction $\boldsymbol{\theta}$ has been chosen with
GDM, a line search is performed to minimize the Lagrangian
(\eqref{E-DM}); that is, $\mathcal{L} (\exp(\ell
\boldsymbol{\theta}))$ is minimized with respect to the step size
$\ell$. Because $\mathcal{L}$ is a quasiperiodic function along $\ell$
\cite{Abrudan2009, Lehtola2013a}, the period of its fastest
oscillation $\lambda_0$ can be estimated from the largest absolute
eigenvalue $|\omega|_\text{max}$ of $\boldsymbol{\theta}$ as
$\lambda_0 = \pi / 2 |\omega|_\text{max}$. This expression can be
derived when the Lagrangian has fourth-order dependence on the
orbitals \cite{Lehtola2016}, as is the case here due to the
two-electron integrals. Relying on earlier experience
\cite{Lehtola2016}, we chose to perform the line search using
parabolic interpolation with a trial step size of $\lambda_0/5$. Once
a optimal value $\ell$ has been found, the orbitals are updated
$\boldsymbol{C} \to \boldsymbol{C} \exp(\ell \boldsymbol{\theta})$ and
the new reference energy $E_{0}$ and density matrices $\gamma$ and
$\Gamma$ are evaluated. Then, the new gradient and diagonal Hessian
are computed, followed by either a new line search, or the
determination that successful convergence has been achieved.

The line search minimization of \eqref{E-DM} has two obvious
approaches. First, one may allow the density matrices $\gamma$ and
$\Gamma$ to relax at every point along the line, which appears the
obvious solution. However, one could also consider \eqref{E-DM} as a
model energy functional by assuming the density matrices $\gamma$ and
$\Gamma$ to be fixed. After some experimentation, we chose to pursue
the latter approach, as in addition to the slight benefit of
eliminating the need to solve new amplitudes during the line search,
it also makes the search direction and preconditioning in GDM
exact. Note that even when the density matrices are frozen within the
line search, every optimization step \emph{must} result in a decrease
of the true CC energy, as the true, relaxed energy will always lie
below the energy predicted by the model functional.

The gradient and diagonal Hessian equations are autogenerated in a
similar fashion to the CC equations, following the machinery detailed
in \citeref{Lehtola2016b}. The density matrices and integrals are
represented as dense subtensors in the molecular orbital (MO) basis,
which is sparse in the PPH. Also the gradient and the diagonal Hessian
are generated in the paired MO basis, as this enabled re-use of the
code developed in \citeref{Lehtola2016b}. However, the Fock matrix
response term in the gradient (the two-electron integral terms in
\eqref{f-grad} in the Appendix) is excluded from the automatic
evaluation, as it would result in the calculation of 3-index molecular
integrals for PP and 4-index molecular orbitals for PQ, which would
dominate the computational cost of the models. Instead, the response
term is evaluated with hand-written code in the atomic orbital (AO)
basis, allowing both for better scaling and for taking advantage of AO
integral sparsity, reducing the cost of evaluating the term to the
same level as Hartree--Fock theory.

The methods have been implemented in a general-purpose library, which
may be made publicly available in the future. The input for a
calculation is composed of the matrix elements, that is the MO basis
Fock and B matrices, which are read in from the disk. The B matrices
can be precomputed using either Cholesky \cite{Beebe1977} or
resolution-of-the-identity \cite{Vahtras1993} techniques. After the
amplitudes have been converged, the computer generated implementation
outputs the energy, the OPDM and the TPDM, as well as the orbital
gradient and the diagonal Hessian as matrices if their calculation has
been requested. This minimal interface guarantees straightforward
interfacing possibilities to various programs.

The results of the present manuscript have been obtained by
interfacing the library with the \textsc{Erkale} program
\cite{Lehtola2012, erkale}. Hartree–Fock occupied orbitals, localized
\cite{Lehtola2013a} with the generalized Pipek–Mezey method
\cite{Lehtola2014} with Becke charges and then paired with
corresponding virtual orbitals using the Sano procedure
\cite{Sano2000}, have been used to initialize the orbital optimization
by GDM. Cholesky decompositions \cite{Beebe1977, Koch2003} are used for all
integral computations, with a $10^{-10}$ screening threshold for the
two-electron integral calculations, and a $10^{-6}$ threshold for the
Cholesky procedure. The AO Cholesky decomposition is run only once for
a given molecule, at the beginning of the Hartree--Fock calculation,
constituting a trivial fraction of the total runtime of the models.

Because single excitations are analogous to orbital rotations
\cite{Thouless1960} and thus imitate orbital relaxation effects within
the active space, the orbital optimization is run without single
excitations as is done in OO-CC and VOO-CC. A gradient convergence
threshold of $\left\Vert \partial \mathcal{L} / \partial \theta_{pq}
\right\Vert \leq 10^{-5} $ is used for the PPH orbital
optimization. Having converged the orbitals with PP or PQ,
single-point calculations are run with PP, PQ, and PH, and density
matrices are calculated. Natural orbitals (NOs) and NO occupation
numbers (NOONs) are obtained by diagonalizing the OPDM.

\section{Results}

While the automated procedures would allow us to formulate an orbital
optimization procedure for all the models in the PPH, in the present
work orbital optimization is undertaken only with PP and PQ, which
have been shown to exhibit excellent scaling with system size
\cite{Lehtola2016b}. Orbital optimization with PH is not considered,
because the truncation of the intermediates that is performed in PH
requires consistent definitions of all the intermediates in the $t$,
$\lambda$ as well as the density matrix equations, but this is not
guaranteed in our code at present. We estimate the effect of the
inconsistent truncation on the NOONs to be negligible (changes in the
occupation numbers of the order of 0.001).

To facilitate comparison with DMRG reference values, the molecular
geometries for the polyacenes were adopted from
\citeref{Hachmann2007}. With the molecule placed on the $xy$ plane,
$\sigma$ and $\pi$ orbital symmetries were assigned by determining the
parity of the orbital $\phi$ upon reflection about the plane
$\phi(x,y,-z)=\pm\phi(x,y,z)$. Like the original Pipek--Mezey method
\cite{Pipek1989}, the generalized Pipek--Mezey method
\cite{Lehtola2014} used in the present work properly separates
$\sigma$ and $\pi$ orbitals, but lacks severe mathematical
shortcomings of the original formulation by Pipek and Mezey that is
based on Mulliken charges \cite{Mulliken1955}.

Orbital optimizations restricted to the $\pi$ and $\sigma$ active
spaces, respectively, were attempted in the STO-3G basis set
\cite{Hehre1969} by deleting the virtual orbitals of the opposite
class. Because STO-3G is a minimal basis set, there are no inactive
virtual orbitals and the optimization problem only needs to be solved
within the active space. For both the $\sigma$- and the $\pi$-space,
the PP optimizations converged within a few dozen iterations. For PQ,
the $\sigma$-space orbital optimizations converged in a few hundred
iterations. However, the convergence of the PQ $\pi$-space
calculations was poor: while for the smaller acenes convergence was
achieved in several hundred iterations, for 6acene the orbital
optimization was still not converged at 500 iterations, and for even
larger acenes the $\lambda$ amplitude equations diverged during the
orbital optimization. (These problems can be understood in light of
the results discussed later on in the manuscript: the restriction to
the $\pi$ space exaggerates correlation effects, which then make
orbital optimization hard.) For this reason, we will not consider
orbital optimization restricted to the $\sigma$ or $\pi$ space further
in the manuscript.

For the full-valence STO-3G orbital optimization with PP, solutions
that break $\sigma$-$\pi$ symmetry by having some orbitals of mixed
character were found for multiple molecules. In contrast, we did not
encounter any broken symmetry solutions with full-valence STO-3G PQ
optimized orbitals. We have thoroughly tested this in the case of
2acene. In analogy to our recent study in self-interaction corrected
density functional theory \cite{Lehtola2016}, 100 different initial
guesses were considered for the initial occupied orbitals, after which
the orbitals were localized with the generalized Pipek--Mezey method
and corresponding virtuals determined with the Sano algorithm. Orbital
optimization initialized with these guesses was performed using both
PP and PQ. For PP, a quarter of the calculations converged on a
symmetry broken solution, while three quarters found the
$\sigma$-$\pi$ preserving solution which is higher in energy. For PQ,
all calculations converged to the same solution that preserves
$\sigma$-$\pi$ symmetry. Calculations were also performed without the
initial generalized Pipek--Mezey / Sano localization procedure, in
which case all PP calculations converged onto the symmetry broken
solution, and PQ again converged to the $\sigma$-$\pi$ preserving
solution.

It is not hard to understand why PP would have more local solutions
than PQ. In PP orbital optimization (i.e. no single excitations), the
only excitations are intra-pair ones: the double excitation $t_{i
  \bar{i}}^{i^\star \bar{i}^\star}$ from the occupied alpha and beta
orbital $i$ and $\bar{i}$, respectively, to the corresponding paired
virtuals $i^{\star}$ and $\bar{i}^{\star}$, respectively. Furthermore,
each amplitude $t_{i \bar{i}}^{i^\star \bar{i}^\star}$ can be solved
analytically, independent of the other pairs \cite{Beran2005}. In
contrast, in PQ orbital optimization, several inter-pair double
excitations $t_{j \bar{i}}^{i^{\star} \bar{i}^{\star}}$, $t_{i
  \bar{j}}^{i^{\star} \bar{i}^{\star}}$, $t_{i \bar{i}}^{j^{\star}
  \bar{i}^{\star}}$, $t_{i \bar{i}}^{i^{\star} \bar{j}^{\star}}$,
$t_{i \bar{i}}^{j^{\star} \bar{j}^{\star}}$, $t_{i \bar{j}}^{i^{\star}
  \bar{j}^{\star}}$, $t_{j \bar{i}}^{j^{\star} \bar{i}^{\star}}$ are
added (as well as same-spin double excitations), as well as
the triple excitations $t_{i j \bar{j}}^{i^{\star} j^{\star}
  \bar{j}^{\star}}$, $t_{i j \bar{j}}^{j^{\small \star} i^{\star}
  \bar{i}^{\star}}$, $t_{j i \bar{i}}^{i^{\star} j^{\star}
  \bar{j}^{\star}}$, and $t_{j i \bar{i}}^{j^{\star} i^{\star}
  \bar{i}^{\small \star}}$ and quadruple excitations $t_{i j \bar{i}
  \bar{j}}^{i^{\star} j^{\star} \bar{i}^{\star}
  \bar{j}^{\star}}$. Thus, PP describes single Kekulé structures,
whereas PQ adds in resonance terms which couple the structures,
eliminating the local solutions. Because the PQ orbitals thus appear
well defined, we chose to use the full-valence optimized PQ orbitals
for all the single-point calculations in our study.

\begin{figure*}
  \subfloat[PP]{
    \begin{centering}
      \includegraphics[width=0.48\textwidth]{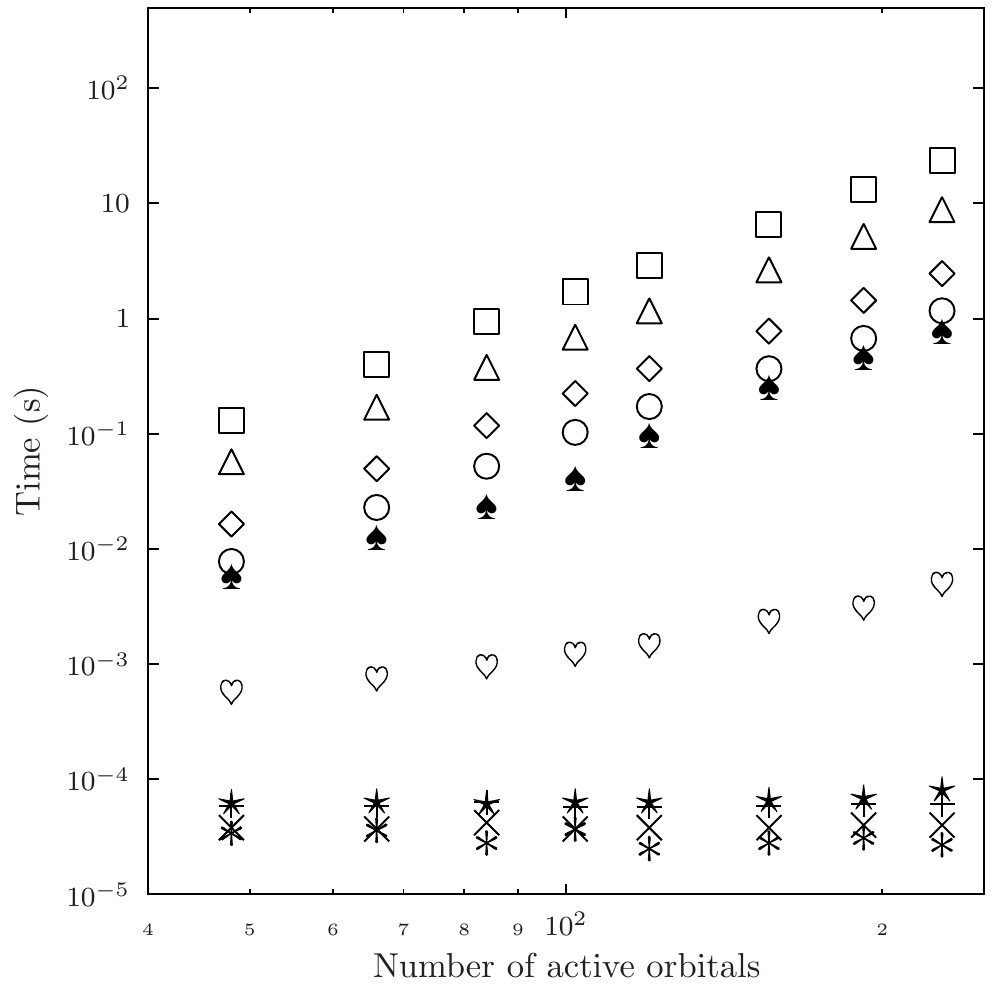}
    \end{centering}
  }
  \subfloat[PQ]{
    \begin{centering}
      \includegraphics[width=0.48\textwidth]{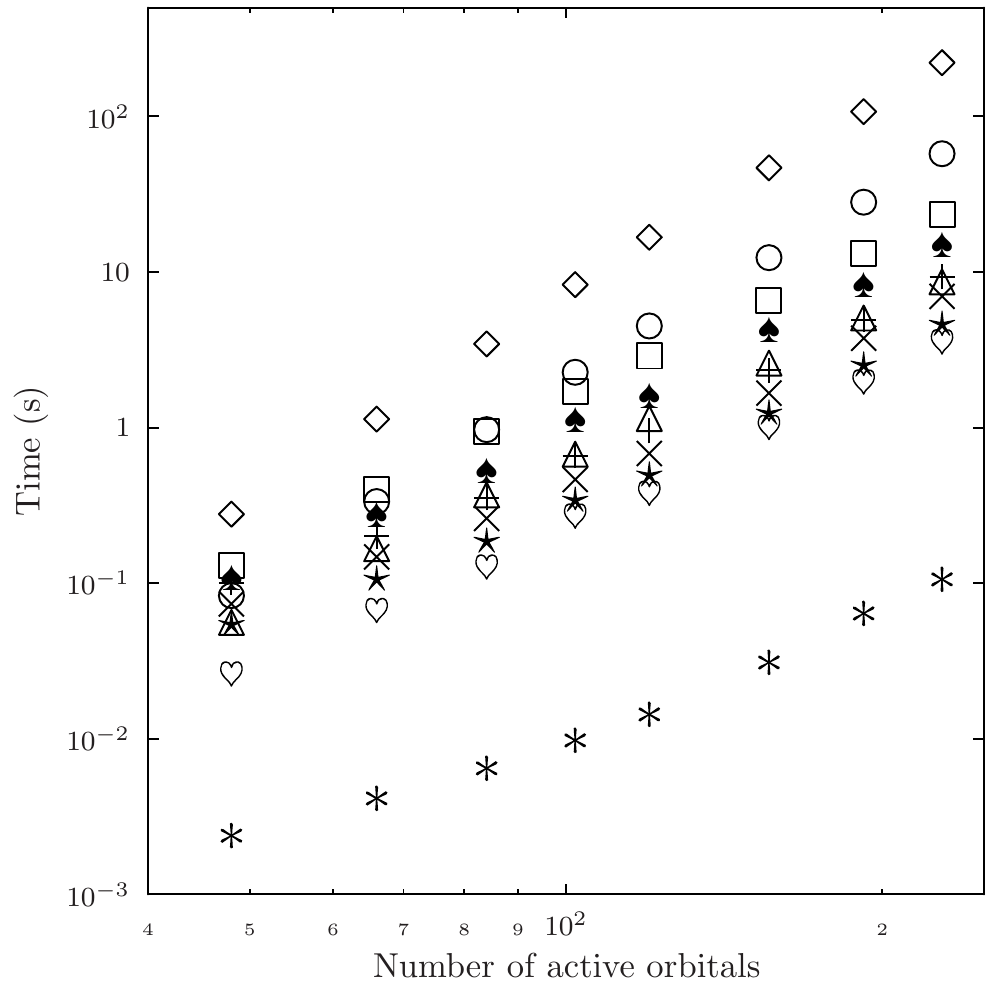}
    \end{centering}
  }
  
  \caption{Computational scaling of the PP and PQ models in the STO-3G
    basis, averaged over 10 iterations. Legend: reference energy $E_0$
    ($\triangle$), $B$ matrix ($\Box$), matrix elements $I$
    ($\bigcirc$), orbital gradient matrix elements $\partial I /
    \partial \theta_{pq}$ ($\diamond$), $t$ amplitude iteration
    ($\times$), $\lambda$ amplitude iteration (+), OPDM $\gamma$ build
    ($\asterisk$), TPDM $\Gamma$ build ($\star$), $\mathcal{L}$
    evaluation ($\heartsuit$), $\partial \mathcal{L} / \partial
    \theta_{pq}$ evaluation ($\spadesuit$).
    \label{fig:scaling}}

\end{figure*}

The scaling of the STO-3G calculations on one of the authors' (S.L.)
workstation, using a single Intel i7-4770 processor core, is shown in
\figref{scaling}. Already in the STO-3G basis, where there are no
inactive virtual orbitals, the integral calculations and the build of
the orbital gradient of the Lagrangian are close to being the rate
determining steps for the orbital optimization. Because everything
else is built only within the active space and thus does not scale
with the size of the basis, the computation of the integrals and the
orbital gradient contraction will become even more dominant in larger
bases: PP has $O(N_\text{act})$ matrix elements and $O(N_\text{act}
N_\text{bf})$ orbital gradient matrix elements, while PQ has
$O(N_\text{act}^2)$ matrix elements and $O(N_\text{act}^2
N_\text{bf})$ orbital gradient matrix elements, where $N_\text{act}$
and $N_\text{bf}$ are the number of active orbitals and basis
functions, respectively, when the Fock response terms are handled
otherwise as discussed above in the Computational Detail section.  An
optimized AO implementation has been reported for PP \cite{Beran2005};
similar techniques might prove beneficial for PQ as well. While PQ
scales asymptotically as $N_\text{act}^4$ \cite{Parkhill2009,
  Lehtola2016b}, PQ is still cubic scaling for the 12acene calculation
with $N_\text{act}=228$, suggesting that enormous calculations will be
possible with a fully optimized implementation.

\subsection{STO-3G basis, $\pi$ space}

To establish the accuracy of the orbital-optimized models, we begin in
the STO-3G basis \cite{Hehre1969} where accurate DMRG data is
available in the $\pi$ space \cite{Hachmann2007}. Calculations are
performed using the $\pi$ orbitals extracted from the full-valence
optimized PQ orbitals. Because $\sigma$-$\pi$ correlation is expected
to be much less important than $\pi$-$\pi$ correlation, the $\pi$
orbitals extracted from the full-valence PQ optimization should not be
exceedingly different from the ones that would be optimal for a
$\pi$-space-only PQ calculation, as all the $\pi$ orbitals of STO-3G
are anyhow included in the full-valence $\pi$ active space.

The correlation energy captured by PP, PQ, and PH, along with the
Hartree--Fock energy are shown in \tabref{Ec-pi-sto3g}.  Because PQ
optimized orbitals are used for all models, results are given both
with and without single excitations included in the single-point
calculation, as these represent relaxations in the active occupied --
active virtual block. (Note however, that the singles do not relax the
other orbital degrees of freedom in PPH, namely, the active occupied
-- active occupied and active virtual -- active virtual rotations.)
One would expect this active space orbital relaxation effect to be
minimal. However, it has also been argued that OO-CC theories that
rely on projective approaches fail to reproduce FCI results
\cite{Kohn2005}, because the singles amplitudes at the optimized
orbitals are not guaranteed to vanish, as they do for true Brueckner
orbitals \cite{Brueckner1955}. Thus, the results obtained including
single excitations should be more accurate than the ones obtained
without singles. But, due to the non-variational character of
projected CC theory, the inclusion of the single excitations may also
result in a significant underestimation of the total energy. The
importance of the inclusion of single excitations is clear from the
results in \tabref{Ec-pi-sto3g}. The energies estimated with single
excitations are noticeably lower than the ones estimated without
single excitations, and are in better agreement with DMRG.

The agreement of PH with DMRG is excellent. Without singles, PH
captures >90\% of the correlation energy for all systems, and with
singles, PH captures >95\% of the correlation energy for all acenes,
even though the $\pi$ electrons have a significant degree of
delocality.

Comparing the numbers of \tabref{Ec-pi-sto3g} with the
analogous data using approximate orbitals (corresponding to the
initial guess used in the present work) in \citeref{Lehtola2016b}
shows remarkable improvement in the PQ results due to the orbital
optimization carried out in the present work. While PQ orbitals are
suboptimal for PP, also the PP energies estimated with PQ orbitals are
significantly lower than those estimated with approximate orbitals.

PQ orbitals are also suboptimal for PH. However, the comparison with
the energies of \citeref{Lehtola2016b} suggests that PQ orbitals yield
lower energies for 2acene--6acene, while for 8acene--12acene the
energies estimated from the approximate orbitals are below the present
estimates obtained with the PQ orbitals. Because PH orbital
optimization has not been attempted in the present work due to
technical issues, it remains to be seen whether this phenomenon is
caused by non-variational effects in PH, or the more delocalized
nature of PH optimal orbitals compared to PQ optimal orbitals as
speculated in \citeref{Lehtola2016b}.

\begin{figure*}
\subfloat[Without singles]{\begin{centering}
\includegraphics[width=0.33\textwidth]{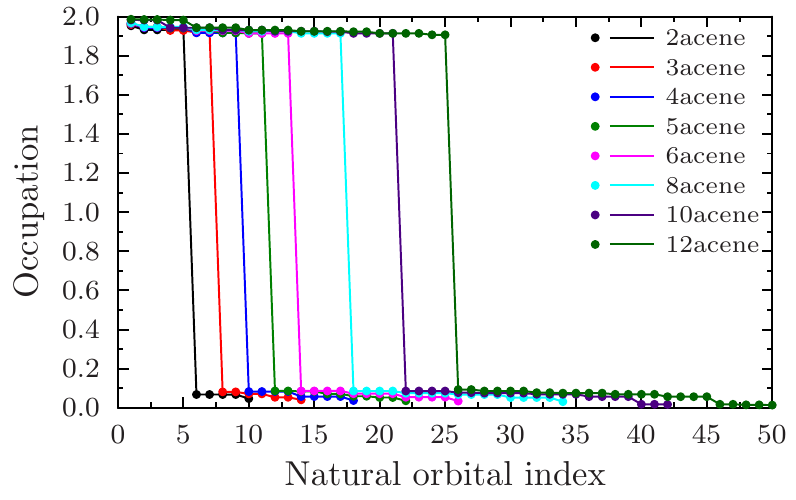}\includegraphics[width=0.33\textwidth]{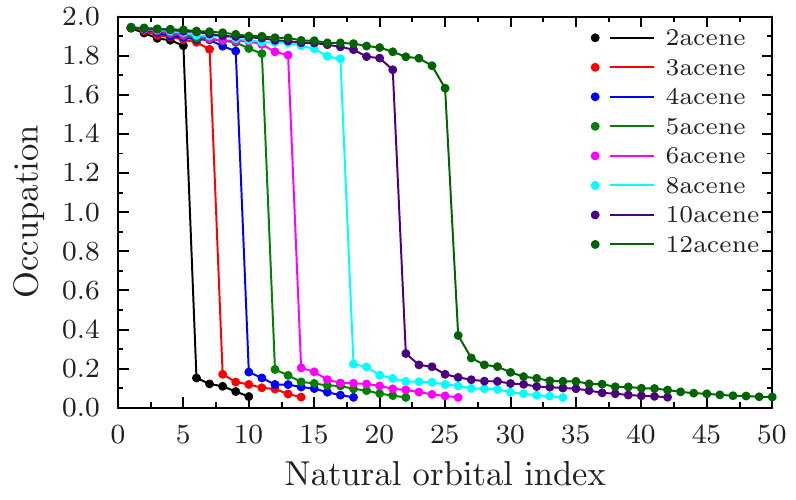}\includegraphics[width=0.33\textwidth]{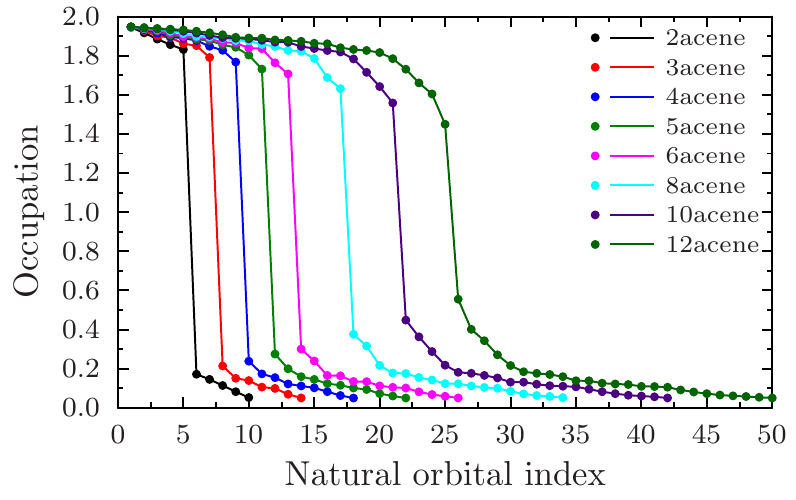}
\par\end{centering}

}

\subfloat[With singles]{\begin{centering}
\includegraphics[width=0.33\textwidth]{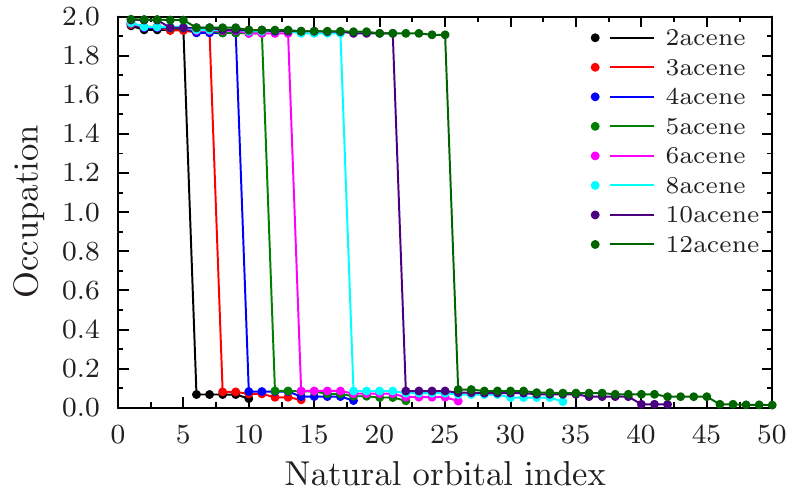}\includegraphics[width=0.33\textwidth]{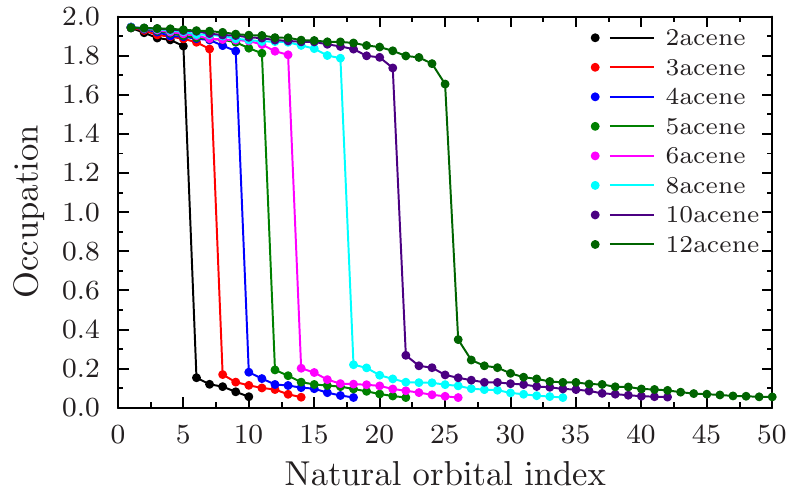}\includegraphics[width=0.33\textwidth]{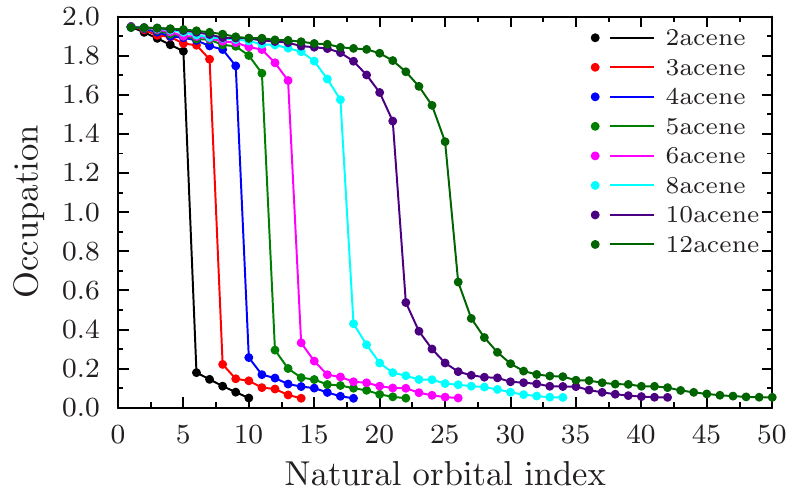}
\par\end{centering}
}

\caption{$\pi$-space NOONs for PP, PQ and PH with and without single
  excitations, using PQ orbitals.\label{fig:pi-noon}}
\end{figure*}

The $\pi$-space NOONs reproduced by PP, PQ, and PH on PQ orbitals are
shown in \figref{pi-noon}. Like the earlier results with approximate
orbitals \cite{Lehtola2016b}, the NOONs reproduced by PH with PQ
orbitals are in qualitative agreement with the DMRG results of
\citeref{Hachmann2007}. However, important differences are noticeable.
Compared to the earlier approximate orbital results
\cite{Lehtola2016b}, the signature of strong correlation is apparent
with optimized orbitals already in PQ. Also, the overcorrelation of PH
for 12acene has been removed, implying it was caused by nonvariational
behavior originating from a poor orbital guess -- this may be like the
difference between nonvariational CCSD energies for bond-stretching
with restricted vs. unrestricted orbitals. The remaining
differences between the PH and DMRG results can be attributed to two
factors: suboptimal orbitals (PQ instead of PH) and the remaining
truncation error in PH.

\begin{sidewaystable*}
  \begin{centering}
    \begin{tabular}{cccccccccc}
      \toprule
      & & &  \multicolumn{3}{c}{Without singles}  & \multicolumn{3}{c}{With singles}  & \tabularnewline
      \cmidrule(l){4-6} \cmidrule(l){7-9}
      Molecule & Active space & $E^\text{HF}$ & PP & PQ & PH &  PP & PQ & PH & DMRG$^a$\tabularnewline
      \midrule
      2acene & (10e,10o) & -378.683899 & -0.115752 & -0.158155 & -0.171489  & -0.115752 & -0.161151 & -0.177136  & -0.178294\tabularnewline
      3acene & (14e,14o) & -529.467421 & -0.159805 & -0.218652 & -0.239966  & -0.159805 & -0.223151 & -0.250444  & -0.254307\tabularnewline
      4acene & (18e,18o) & -680.246082 & -0.210661 & -0.284669 & -0.312433  & -0.210661 & -0.290035 & -0.325936  & -0.332661\tabularnewline
      5acene & (22e,22o) & -831.022093 & -0.256128 & -0.347849 & -0.383498  & -0.256128 & -0.354469 & -0.402041  & -0.412627\tabularnewline
      6acene & (26e,26o) & -981.794501 & -0.300955 & -0.412119 & -0.456519  & -0.300955 & -0.420058 & -0.480374  & -0.495683\tabularnewline
      8acene & (34e,34o) & -1283.332597 & -0.374849 & -0.531322 & -0.599087  & -0.374849 & -0.542803 & -0.639394 & -0.668526\tabularnewline
      10acene & (42e,42o) & -1584.874961 & -0.430551 & -0.668290 & -0.752407  & -0.430552 & -0.681619 & -0.797103  & -0.838580\tabularnewline
      12acene & (50e,50o) & -1886.418502 & -0.488271 & -0.806117 & -0.909540  & -0.488271 & -0.822062 & -0.958267  & -1.007378$^{b}$\tabularnewline
      \bottomrule
    \end{tabular}
    \par\end{centering}
    \tabnote{ $^a$Data from \citeref{Hachmann2007}. \\ $^{b}$The DMRG
      reference value of \citeref{Hachmann2007} may not be converged.  }
    \caption{$\pi$-space STO-3G correlation energy captured by the models
      using PQ orbitals, without and with single excitations, compared to
      DMRG reference data. The Hartree--Fock
      energy $E^\text{HF}$ is also given.
      \label{tab:Ec-pi-sto3g}}
\end{sidewaystable*}

\subsection{STO-3G basis, full-valence calculations}

Having discussed the accuracy of the models in the $\pi$ space, we are
in a good position to continue onto the full valence space. The
excellent accuracy in the $\pi$ space is noteworthy, because as
discussed in \citeref{Lehtola2016b}, the perfect pairing hierarchy can
be seen as a local correlation model. As the models are able to
successfully treat the delocalized $\pi$ electrons in the polyacenes,
they will be even better in treating the additional correlation due to
the $\sigma$ electrons, which are localized.

\begin{figure*}
  \subfloat[Without singles]{\begin{centering}
      \includegraphics[width=0.33\textwidth]{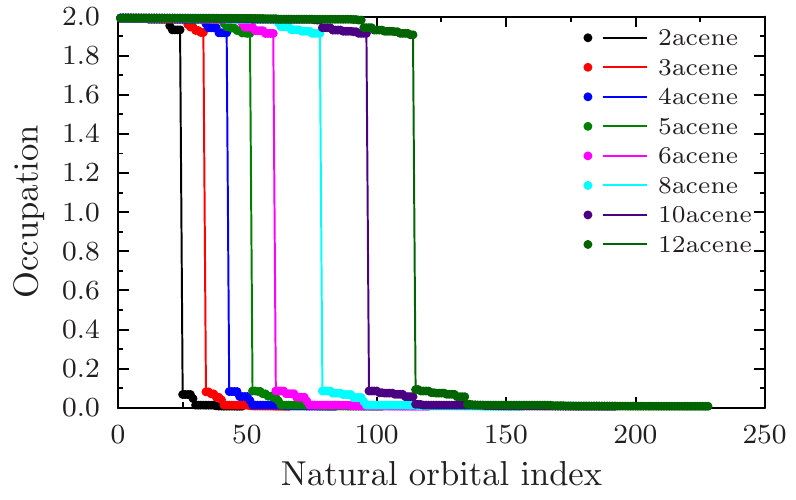}\includegraphics[width=0.33\textwidth]{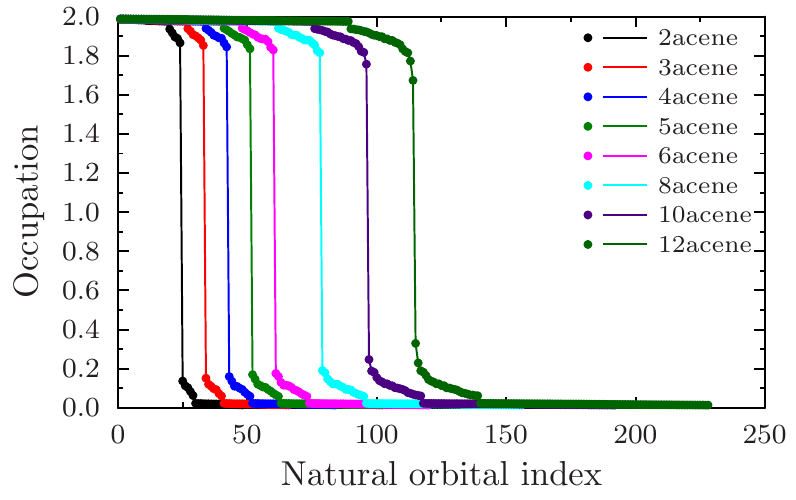}\includegraphics[width=0.33\textwidth]{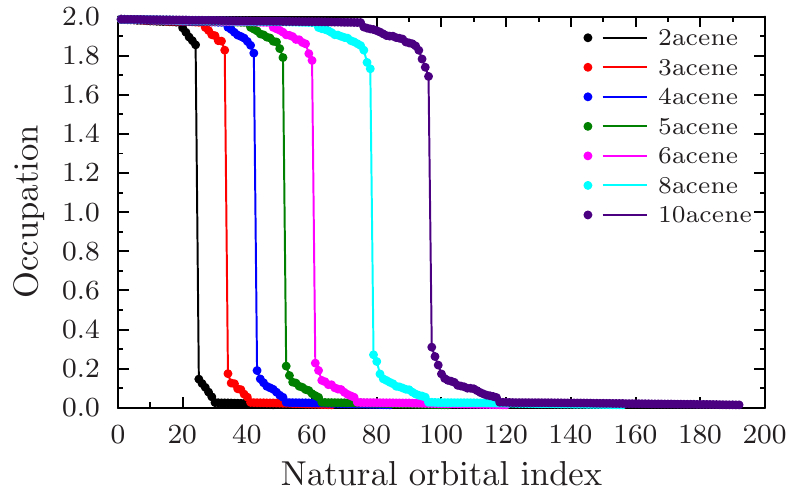}
      \par\end{centering}
  }

  \subfloat[With singles]{\begin{centering}
      \includegraphics[width=0.33\textwidth]{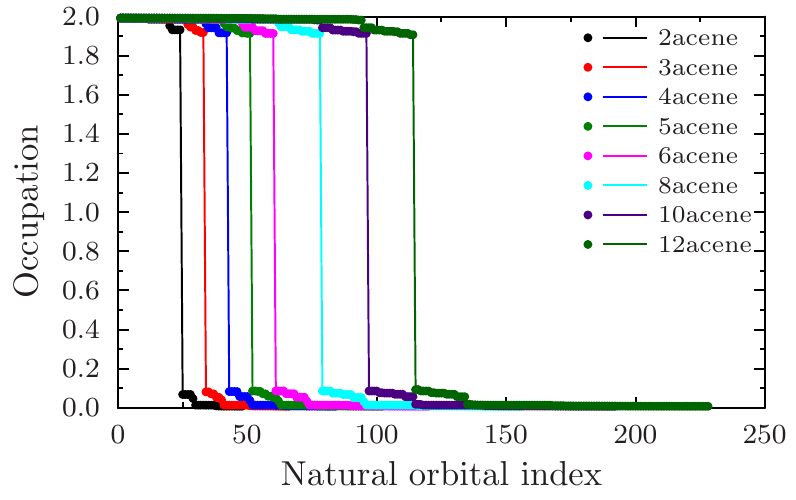}\includegraphics[width=0.33\textwidth]{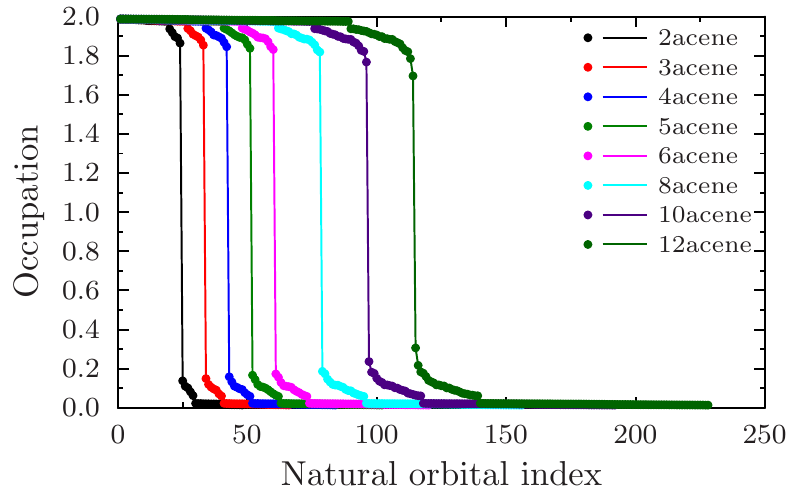}\includegraphics[width=0.33\textwidth]{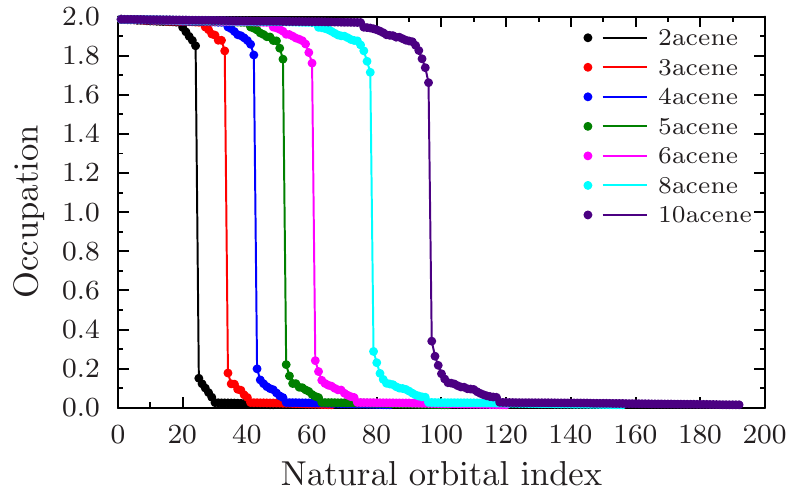}
      \par\end{centering}
  }

  \caption{Full-valence STO-3G NOONs for PP, PQ and PH with
    and without single excitations, using PQ orbitals.\label{fig:noon-fv-sto3g}}
\end{figure*}

The results of our full-valence STO-3G calculations are shown in
\tabref{Ec-fv-sto3g} for the correlation energies, and
\figref{noon-fv-sto3g} for the NOONs. As is evident from the plot, the
strong correlation that is clear in the $\pi$ space calculation is
quenched in the full-valence calculation. \Tabref{honogap-fv-sto3g}
shows the gap between the highest occupied NO (HONO) and the lowest
unoccupied NO (LUNO) in the full-valence calculation, which is
significantly larger than the one in the $\pi$-space only calculation
in \tabref{honogap-pi-sto3g}.

The single excitations are again found to play a significant
role. Contrary to the previous case dealing with the subset of $\pi$
orbitals, here the orbitals have been optimized \emph{for the same
  problem}. Thus, if the orbitals had a quasi-Brueckner character,
there would be little difference between the results obtained with and
without single excitations. In contrast, the results in
\tabref{Ec-fv-sto3g} clearly show that this is not the case: single
excitations are necessary even when the orbitals have been optimized.

\begin{sidewaystable*}
  \begin{centering}
    \begin{tabular}{ccccccccc}
      \toprule
      & & & \multicolumn{3}{c}{Without singles}  & \multicolumn{3}{c}{With singles} \tabularnewline
      \cmidrule(l){4-6} \cmidrule(l){7-9}
      Molecule & Active space & Hilbert space & PP & PQ & PH  &  PP & PQ & PH\tabularnewline
      \midrule
      2acene & (48e,48o) & $1.0\times10^{27}$ & -0.322737 & -0.577926 & -0.681109 & -0.322737 & -0.582018 & -0.688946\tabularnewline
      3acene & (66e,66o) & $5.2\times10^{37}$ & -0.440449 & -0.795324 & -0.945119 & -0.440449 & -0.801436 & -0.958952\tabularnewline
      4acene & (84e,84o) & $2.8\times10^{48}$ & -0.564961 & -1.019499 & -1.213242 & -0.564962 & -1.026913 & -1.231753\tabularnewline
      5acene & (102e,102o) & $1.6\times10^{59}$ & -0.684079 & -1.240375 & -1.479518 & -0.684080 & -1.249531 & -1.504362\tabularnewline
      6acene & (120e,120o) & $9.3\times10^{69}$ & -0.802603 & -1.462114 & -1.747772 & -0.802604 & -1.473113 & -1.779061\tabularnewline
      8acene & (156e,156o) & $3.4\times10^{91}$ & -1.023950 & -1.894352 & -2.279430 & -1.023952 & -1.910179 & -2.327852\tabularnewline
      10acene & (192e,192o) & $1.3\times10^{113}$ & -1.227908 & -2.323432 & -2.819420& -1.227910 & -2.341611 & -2.872236\tabularnewline
      12acene & (228e,228o) & $5.2\times10^{134}$ & -1.433817 & -2.750959 &  & -1.433820 & -2.772657 & \tabularnewline
      \bottomrule
    \end{tabular}
    \par\end{centering}
    \caption{Correlation energies in the STO-3G full valence space.\label{tab:Ec-fv-sto3g}}
\end{sidewaystable*}

\begin{table*}
  \begin{centering}
    \begin{tabular}{ccccccc}
      \toprule
      & \multicolumn{3}{c}{Without singles}  & \multicolumn{3}{c}{With singles} \tabularnewline
      \cmidrule(l){2-4} \cmidrule(l){5-7}
      \midrule
      2acene & 1.8642 & 1.7288 & 1.7091 & 1.8642 & 1.7266 & 1.6998\tabularnewline
      3acene & 1.8372 & 1.7011 & 1.6538 & 1.8372 & 1.7046 & 1.6469\tabularnewline
      4acene & 1.8340 & 1.6852 & 1.6227 & 1.8340 & 1.6872 & 1.6045\tabularnewline
      5acene & 1.8280 & 1.6658 & 1.5775 & 1.8280 & 1.6710 & 1.5616\tabularnewline
      6acene & 1.8280 & 1.6535 & 1.5474 & 1.8280 & 1.6585 & 1.5211\tabularnewline
      8acene & 1.8286 & 1.6242 & 1.4624 & 1.8286 & 1.6326 & 1.4271\tabularnewline
      10acene & 1.8286 & 1.5095 & 1.3842 & 1.8286 & 1.5311 & 1.3216\tabularnewline
      12acene & 1.8138 & 1.3446 &  & 1.8138 & 1.3905 & \tabularnewline
      \bottomrule
    \end{tabular}
    \par\end{centering}
    \caption{Full-valence STO-3G HONO-LUNO gaps for PP, PQ, PH.\label{tab:honogap-fv-sto3g}}
\end{table*}

\begin{table*}
  \begin{centering}
    \begin{tabular}{cccccccc}
      \toprule
      & \multicolumn{3}{c}{Without singles}  & \multicolumn{3}{c}{With singles} & \tabularnewline
      \cmidrule(l){2-4} \cmidrule(l){5-7}
      Molecule & PP & PQ & PH  & PP & PQ & PH & DMRG$^a$\tabularnewline
      \midrule
      2acene & 1.8642 & 1.6990 & 1.6600 & 1.8642 & 1.6956 & 1.6433 & 1.6315\tabularnewline
      3acene & 1.8372 & 1.6617 & 1.5770 & 1.8372 & 1.6645 & 1.5602 & 1.5303\tabularnewline
      4acene & 1.8340 & 1.6411 & 1.5293 & 1.8340 & 1.6418 & 1.4913 & 1.4213\tabularnewline
      5acene & 1.8280 & 1.6150 & 1.4570 & 1.8280 & 1.6190 & 1.4154 & 1.3063\tabularnewline
      6acene & 1.8280 & 1.5992 & 1.4066 & 1.8280 & 1.6025 & 1.3409 & 1.1567\tabularnewline
      8acene & 1.8286 & 1.5608 & 1.2559 & 1.8286 & 1.5671 & 1.1450 & 0.7885\tabularnewline
      10acene & 1.8286 & 1.4507 & 1.1102 & 1.8286 & 1.4692 & 0.9276 & 0.5777\tabularnewline
      12acene & 1.8138 & 1.2641 & 0.8942 & 1.8138 & 1.3063 & 0.7175 & 0.4717$^b$\tabularnewline
      \bottomrule
    \end{tabular}
    \par\end{centering}
    \tabnote{ $^a$Data from \citeref{Hachmann2007}. \\ $^{b}$The DMRG
      reference value of \citeref{Hachmann2007} may not be converged.  }
    \caption{$\pi$-space HONO-LUNO gaps for PP, PQ, PH, compared with DMRG data
      from \citeref{Hachmann2007}.\label{tab:honogap-pi-sto3g}}
\end{table*}

\subsection{cc-pVDZ basis}

\begin{figure*}
\subfloat[Without singles]{\begin{centering}
\includegraphics[width=0.33\textwidth]{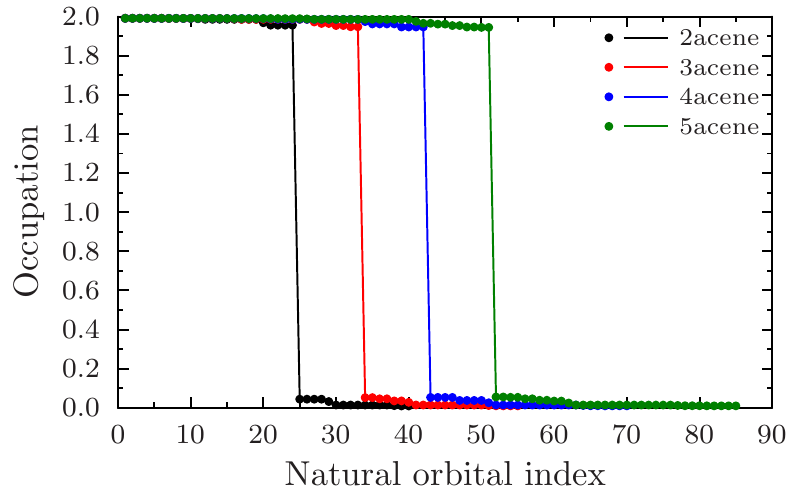}\includegraphics[width=0.33\textwidth]{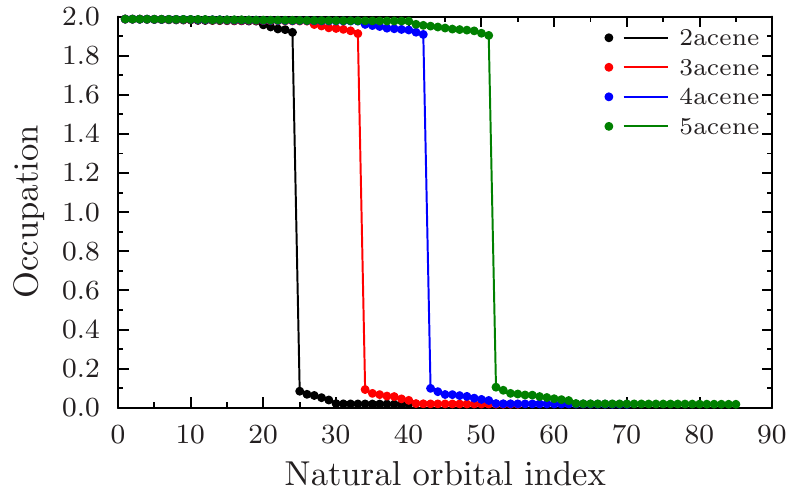}\includegraphics[width=0.33\textwidth]{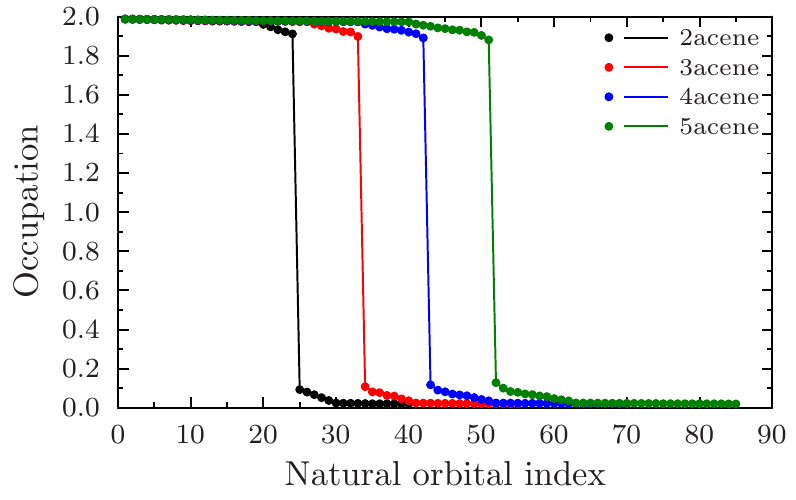}
\par\end{centering}
}

\subfloat[With singles]{\begin{centering}
\includegraphics[width=0.33\textwidth]{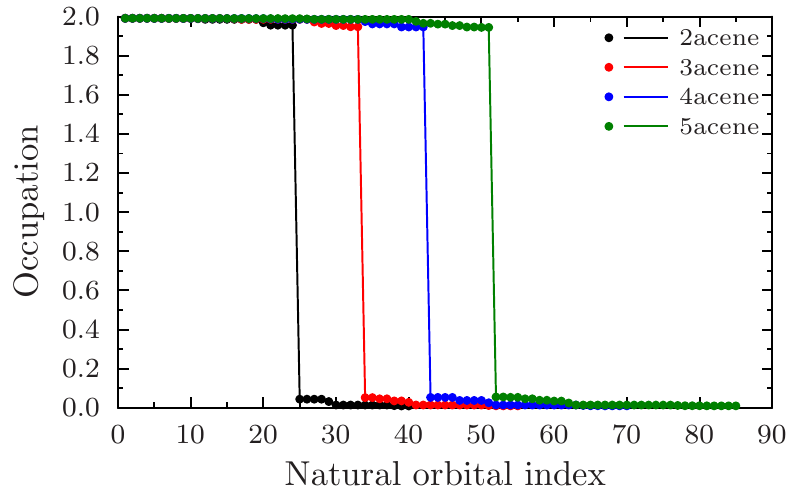}\includegraphics[width=0.33\textwidth]{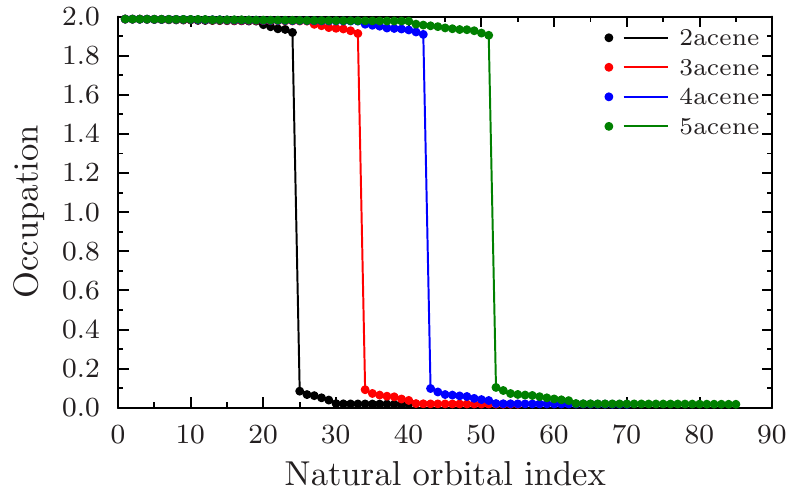}\includegraphics[width=0.33\textwidth]{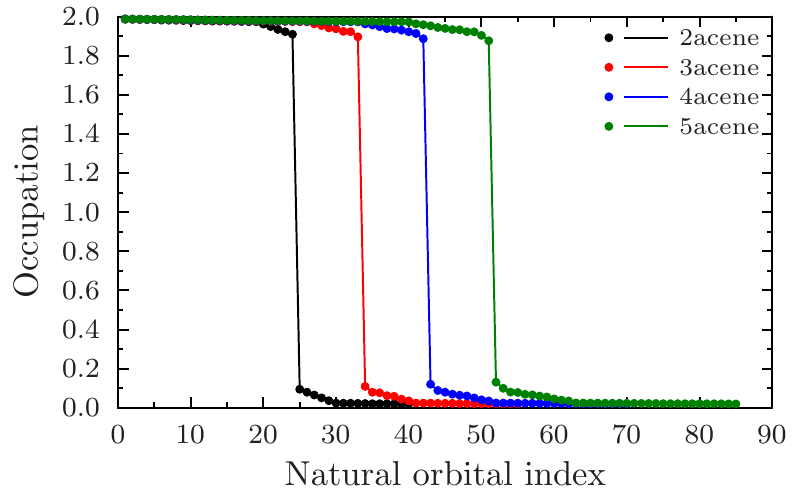}
\par\end{centering}
}

\caption{Full-valence cc-pVDZ NOONs for PP, PQ and PH
with and without single excitations, using PQ orbitals.\label{fig:noon-fv-ccpvdz}}
\end{figure*}

Full-valence orbital optimizations were performed up to 5acene in the
cc-pVDZ basis set \cite{Dunning1989}. The resulting correlation
energies, NOONs, and HONO-LUNO gaps are shown in
\tabref{Ec-fv-ccpvdz}, \figref{noon-fv-ccpvdz}, and
\tabref{honogap-fv-ccpvdz}, respectively. The correlation energies are
smaller in cc-pVDZ than in STO-3G, and the HONO-LUNO gaps are
larger. While the strong correlation of the $\pi$ electrons was
already quenched in the full valence space of STO-3G, the signature of
strong correlation is even more strongly quenched in the cc-pVDZ
basis, even though the size of the used active space is the same in
STO-3G and cc-pVDZ. As a result of the smaller amount of correlation,
the PP, PQ and PH results are closer to each other in cc-pVDZ than in
STO-3G. It is thus clear that the limited flexibility of the STO-3G
basis set results in an exaggeration of the correlation effects,
and the question about the polyradicaloid character of the polyacenes
may still be unresolved.

\begin{sidewaystable*}
\begin{centering}
    \begin{tabular}{cccccccccc}
      \toprule
      & & & \multicolumn{3}{c}{Without singles}  & \multicolumn{3}{c}{With singles} \tabularnewline
      \cmidrule(l){4-6} \cmidrule(l){7-9}
      Molecule & Active space & $E^\text{HF}$ & PP & PQ & PH  & PP & PQ & PH\tabularnewline
      \midrule
      2acene & (48e,48o) & -383.383836 & -0.301755 & -0.533474 & -0.612158 & -0.301756 & -0.536128 & -0.617254\tabularnewline
      3acene & (66e,66o) & -536.037636 & -0.412436 & -0.734416 & -0.848505 & -0.412438 & -0.738267 & -0.857082\tabularnewline
      4acene & (84e,84o) & -688.687483 & -0.527842 & -0.940703 & -1.088061 & -0.527846 & -0.945346 & -1.099478\tabularnewline
      5acene & (102e,102o) & -841.335342 & -0.639575 & -1.144255 & -1.325955 & -0.639581 & -1.149943 & -1.340944\tabularnewline
      \bottomrule
    \end{tabular}
    \par\end{centering}
    \caption{Correlation energies in the full cc-pVDZ valence active
      space. The Hartree--Fock energy $E^\text{HF}$ is also given
      .\label{tab:Ec-fv-ccpvdz}}
\end{sidewaystable*}

\begin{table*}
  \begin{centering}
    \begin{tabular}{cccccccc}
      \toprule
      & \multicolumn{3}{c}{Without singles}  & \multicolumn{3}{c}{With singles} \tabularnewline
      \cmidrule(l){2-4} \cmidrule(l){5-7}
      Molecule & PP & PQ & PH  & PP & PQ & PH\tabularnewline
      \midrule
      2acene & 1.9120 & 1.8348 & 1.8185 & 1.9120 & 1.8340 & 1.8149\tabularnewline
      3acene & 1.8958 & 1.8198 & 1.7907 & 1.8958 & 1.8212 & 1.7871\tabularnewline
      4acene & 1.8939 & 1.8094 & 1.7737 & 1.8939 & 1.8102 & 1.7665\tabularnewline
      5acene & 1.8894 & 1.7984 & 1.7525 & 1.8894 & 1.8004 & 1.7454\tabularnewline
      \bottomrule
    \end{tabular}
    \par\end{centering}
    \caption{Full-valence cc-pVDZ HONO-LUNO gaps for PP, PQ, PH.\label{tab:honogap-fv-ccpvdz}}
\end{table*}

\section{Summary and Discussion}

We have presented the efficient implementation of orbital optimization
for the models in the perfect pairing hierarchy, and applied the new
implementation to calculations on linear polyacenes. Our results show
that orbital optimization is feasible for PQ for huge system sizes –
up to (228e,228o) in the present work – and that the optimization
significantly improves the quality of the results. The cost of orbital
optimization in PP and PQ is dominated by the evaluation of matrix
elements, the cost of which can be reduced by various optimization
techniques, such as evaluation in the atomic orbital basis.

While it was found that PP may produce broken symmetry solutions that
lie energetically below the $\sigma$-$\pi$ symmetry preserving
solution, only a single PQ solution -- which furthermore preserves
$\sigma$-$\pi$ symmetry -- was found for every molecule in the
polyacene series. For this reason, full-valence optimized PQ orbitals
were used for all the single-point calculations in the present work.

As has been argued in the literature \cite{Kohn2005}, even when
optimized orbitals are used, it is necessary to include single
excitations in the treatment, because the Brueckner condition
\cite{Brueckner1955} may not be satisfied in projected CC theories
which are non-variational. We have demonstrated that this is true also
within the PPH: the inclusion of single excitations leads to
significant energy lowerings in full-valence PQ calculations using
optimized orbitals. In principle this energy lowering may also be due
to non-variational effects, but we did not witness any breaking of
variationality in the present work.

We have shown that PH used on top of PQ orbitals successfully captures
the strong correlation in the $\pi$ space of the polyacenes in STO-3G,
capturing over 95\% of the correlation energy relative to DMRG. While
PH is already successful in the delocalized $\pi$ space, because it can
be understood as a three-pair local correlation model
\cite{Lehtola2016b}, it should perform even better for the additional
description of the localized $\sigma$ electrons. We have shown that
once the $\sigma$ electrons are included to complete the full valence
active space, the strong correlation of the $\pi$ electrons of the
polyacenes is quenched in the STO-3G calculations. A similar result
has been obtained contemporaneously in our group through coupled
cluster valence bond calculations \cite{Lee2017}. The reason why the
results of \citeref{Lee2017} agree with ours can be understood by
analysis of the results of the PH calculations. In the method of
\citeref{Lee2017}, excitations are included to the approximate
quadruples level. In our PH calculations, the role of quintuple and
hextuple excitations (which CCVB lacks) is small, indicating they
describe dynamic correlation effects that have negligible importance
on the natural orbital occupation numbers.

We have further presented the first accurate ab initio calculations on
polyacenes that go beyond the minimal STO-3G basis set. Our
calculations in the cc-pVDZ basis set, employing the same active
spaces as in STO-3G, show that the adoption of a non-minimal basis set
further reduces the signature of strong correlation in the full
valence space. Thus, the polyradicaloid character of the smaller
acenes claimed in works by multiple authors may then be purely an
effect of an insufficient active space and basis set, highlighting the
need for further calculations, as the onset of polyradicaloid behavior
in the acene series may be delayed when all valence electrons are
correlated.

Orbital optimization for strong correlation problems is often
non-trivial due to the existence of broad minima. For this reason,
many approaches that guarantee second-order convergence have been
developed for MC-SCF \cite{Lengsfield1980, Lengsfield1981,
  Siegbahn1981, Werner1981, Yeager1982, Jorgensen1983b, Jensen1984,
  Werner1985, Knowles1985, Jensen1987, Hoffmann2002} and DMRG
\cite{Ghosh2008, Ma2016, Sun2017} calculations. While the Lagrangian
formulation (\eqref{E-DM}) is not correct to second order in the
orbitals, it has been recently used for a quadratic convergence OO-CC
implementation \cite{Bozkaya2011}. A similar approach should be
feasible for the PPH as well -- possibly supplemented by the use of
the proxy function approach (freezing the density matrices between
orbital update steps) which has been used in the present
work. Although the cost per iteration as well as storage requirements
in such an approach would be significantly higher than in the present
one based on GDM, a second-order approach may yield converged orbitals
in considerably fewer iterations and could be pursued in future work.

\section*{Funding}

This work was supported by the Director, Office of Basic Energy Sciences,
Chemical Sciences, Geosciences, and Biosciences Division of the U.S.
Department of Energy, under Contract No. DE-AC02-05CH11231.

\section*{Appendix}

Here, we give the equations necessary to implement the rotation gradients
and diagonal Hessians of the integrals. The rotation matrix $\boldsymbol{U}$
that is used to rotate the orbitals $\boldsymbol{C}$
\begin{align*}
\boldsymbol{C}\to & \boldsymbol{C}\boldsymbol{U}
\end{align*}
is parametrized as
\begin{align*}
\boldsymbol{U}= & \exp\boldsymbol{\theta}
\end{align*}
where $\boldsymbol{\theta}$ is an anti-hermitian matrix of rotation
parameters. Due to the antihermicity, we can parametrize the rotations
with either only the upper or only the lower triangular part of \textbf{$\boldsymbol{\theta}$}.
Using the former choice we write
\begin{align*}
\boldsymbol{\theta}= & \left(\begin{array}{ccccc}
0 & \Delta_{12} & \Delta_{13} & \cdots & \Delta_{1n}\\
-\Delta_{12} & 0 & \Delta_{23} & \cdots & \Delta_{2n}\\
-\Delta_{13} & -\Delta_{23} & 0 & \cdots & \Delta_{3n}\\
\vdots & \vdots & \vdots & \ddots & \vdots\\
-\Delta_{1n} & -\Delta_{2n} & -\Delta_{3n} & \cdots & 0
\end{array}\right).
\end{align*}
or more symbolically
\begin{align*}
\boldsymbol{\theta}= & \left(\begin{array}{cc}
0 & \boldsymbol{\Delta}\\
-\boldsymbol{\Delta}^{\text{T}} & 0
\end{array}\right).
\end{align*}
The derivative with respect to the irreducible parameters $\boldsymbol{\Delta}$
is obtained as
\begin{align}
\frac{\partial E}{\partial\Delta_{tu}}= & \sum_{rs}\frac{\partial\theta_{rs}}{\partial\Delta_{tu}}\frac{\partial E}{\partial\theta_{rs}}\nonumber \\
= & \frac{\partial E}{\partial\theta_{tu}}-\frac{\partial E}{\partial\theta_{ut}}\label{eq:Egrad}
\end{align}
and the diagonal Hessian as

\begin{align}
\frac{\partial^{2}E}{\partial\Delta_{tu}^{2}}= & \sum_{pqrs}\frac{\partial\theta_{pq}}{\partial\Delta_{tu}}\frac{\partial}{\partial\theta_{pq}}\left(\frac{\partial\theta_{rs}}{\partial\Delta_{tu}}\frac{\partial E}{\partial\theta_{rs}}\right)\nonumber \\
= & \sum_{pqrs}\left(\delta_{pt}\delta_{qu}-\delta_{pu}\delta_{qt}\right)\left(\delta_{rt}\delta_{su}-\delta_{ru}\delta_{st}\right)\frac{\partial^{2}E}{\partial\theta_{pq}\partial\theta_{rs}}\nonumber \\
= & \frac{\partial^{2}E}{\partial\theta_{tu}\partial\theta_{tu}}-2\frac{\partial^{2}E}{\partial\theta_{tu}\partial\theta_{ut}}+\frac{\partial^{2}E}{\partial\theta_{ut}\partial\theta_{ut}}.\label{eq:Ehess}
\end{align}

The starting point to obtain the rotation gradients and diagonal Hessians
is to write integrals rotated by an angle $\boldsymbol{\theta}$ as
\begin{align}
h_{pq}= & U_{Pp}U_{Qq}h_{PQ},\label{eq:h-pq}\\
v_{pqrs}= & U_{Pp}U_{Qq}U_{Rr}U_{Ss}v_{PQRS},\label{eq:v-pqrs}
\end{align}
where the capital indices $P$, $Q$, $R$ and $S$ refer to the reference
set of orbitals, $h_{pq}$ are one-electron integrals, and the shorthand
notation $v_{pqrs}=\braket{pq||rs}$ is used for the two-electron
integrals. Then, by substituting the Taylor expansion
\begin{align}
\boldsymbol{U}= & \exp\boldsymbol{\boldsymbol{\theta}}\approx1+\boldsymbol{\theta}+\frac{1}{2}\boldsymbol{\theta}^{2}\label{eq:U-taylor}
\end{align}
the gradient with respect to the rotation parameters $\boldsymbol{\theta}$
can be calculated, and evaluated at the reference orbitals ($\boldsymbol{\theta}=\boldsymbol{0}$)
to give the orbital rotation gradient

\begin{align}
\left(\frac{\partial h_{pq}}{\partial\theta_{tu}}\right)_{\boldsymbol{\theta}=\boldsymbol{0}}= & h_{tq}\delta_{pu}+h_{pt}\delta_{qu},\label{eq:h-grad}
\end{align}
where $\delta_{pq}$ is the Kronecker delta symbol. Similarly, the
gradient for the two-electron integral is
\begin{align}
\left(\frac{\partial v_{pqrs}}{\partial\theta_{tu}}\right)_{\boldsymbol{\theta}=\boldsymbol{0}}= & v_{tqrs}\delta_{pu}+v_{ptrs}\delta_{qu}+v_{pqts}\delta_{ru}+v_{pqrt}\delta_{su}.\label{eq:v-grad}
\end{align}
The Fock matrix is given by
\begin{align}
f_{pq}= & h_{pq}+\sum_{k\in\text{occ.}}v_{pkqk},\label{eq:fock}
\end{align}
where the sum over $k$ runs over occupied spin-orbitals, and thus
the gradient from the Fock matrix elements also gathers a two-electron
response term
\begin{align}
\left(\frac{\partial f_{pq}}{\partial\theta_{tu}}\right)_{\boldsymbol{\theta}=\boldsymbol{0}}=\left(\frac{\partial f_{pq}^{0}}{\partial\theta_{tu}}\right)_{\boldsymbol{\theta}=\boldsymbol{0}}+ & \left(v_{ptqu}+v_{puqt}\right)\delta_{u,\text{occ.}}.\label{eq:f-grad}
\end{align}
Here, $(\partial f_{pq}^{0}/\partial_{tu})_{\boldsymbol{\theta}=\boldsymbol{0}}$
denotes the gradient of the ``frozen'' Fock matrix, which is given
by \eqref{h-grad}. Similar expressions to \eqrangeref{h-grad}{f-grad}
have been presented in \citeref{Sherrill1998}. Using the same methodology,
the diagonal Hessian components can be found out to be

\begin{align}
\left(\frac{\partial^{2}h_{pq}}{\partial\theta_{tu}^{2}}\right)_{\boldsymbol{\theta}=\boldsymbol{0}}= & h_{tq}\delta_{tu}\delta_{pt}+2h_{tt}\delta_{pu}\delta_{qu}+h_{pt}\delta_{tu}\delta_{qt},\label{eq:h-direct}\\
\left(\frac{\partial^{2}h_{pq}}{\partial\theta_{tu}\partial\theta_{ut}}\right)_{\boldsymbol{\theta}=\boldsymbol{0}}= & \frac{1}{2}h_{tq}\delta_{pt}+\frac{1}{2}h_{uq}\delta_{pu}+\frac{1}{2}h_{pt}\delta_{qt}+\frac{1}{2}h_{pu}\delta_{qu}\label{eq:h-cross}\\
+ & h_{tu}\delta_{pu}\delta_{qt}+h_{ut}\delta_{pt}\delta_{qu}\nonumber \\
\left(\frac{\partial^{2}f_{pq}}{\partial\theta_{tu}^{2}}\right)_{\boldsymbol{\theta}=\boldsymbol{0}}=\left(\frac{\partial^{2}f_{pq}^{0}}{\partial\theta_{tu}^{2}}\right)_{\boldsymbol{\theta}=\boldsymbol{0}}+ & \left[2(1+\delta_{tu})v_{ptqt}+2v_{tuqt}\delta_{pu}+2v_{pttu}\delta_{qu}\right]\delta_{u,\text{occ.}}\label{eq:f-direct}\\
\left(\frac{\partial^{2}f_{pq}}{\partial\theta_{tu}\partial\theta_{ut}}\right)_{\boldsymbol{\theta}=\boldsymbol{0}}=\left(\frac{\partial^{2}f_{pq}^{0}}{\partial\theta_{tu}\partial\theta_{ut}}\right)_{\boldsymbol{\theta}=\boldsymbol{0}}+ & v_{ptqt}\left(1+\delta_{tu}\right)\delta_{t,\text{occ.}}+v_{tuqt}\delta_{pu}\delta_{t,\text{occ.}}+v_{pttu}\delta_{qu}\delta_{t,\text{occ.}}\label{eq:f-cross}\\
+ & v_{puqu}\left(1+\delta_{tu}\right)\delta_{u,\text{occ.}}+v_{utqu}\delta_{pt}\delta_{u,\text{occ.}}+v_{puut}\delta_{qt}\delta_{u,\text{occ}.}\nonumber \\
\left(\frac{\partial^{2}v_{pqrs}}{\partial\theta_{tu}^{2}}\right)_{\boldsymbol{\theta}=\boldsymbol{0}}= & v_{tqrs}\delta_{tu}\delta_{pt}+v_{ptrs}\delta_{tu}\delta_{qt}+v_{pqts}\delta_{tu}\delta_{rt}+v_{pqrt}\delta_{tu}\delta_{st}\label{eq:v-direct}\\
+ & 2v_{tqts}\delta_{pu}\delta_{ru}+2v_{tqrt}\delta_{pu}\delta_{su}+2v_{ptts}\delta_{qu}\delta_{ru}+2v_{ptrt}\delta_{qu}\delta_{su}\nonumber \\
\left(\frac{\partial^{2}v_{pqrs}}{\partial\theta_{tu}\partial\theta_{ut}}\right)_{\boldsymbol{\theta}=\boldsymbol{0}}= & \frac{1}{2}v_{tqrs}\delta_{pt}+\frac{1}{2}v_{uqrs}\delta_{pu}+\frac{1}{2}v_{ptrs}\delta_{qt}+\frac{1}{2}v_{purs}\delta_{qu}\label{eq:v-cross}\\
+ & \frac{1}{2}v_{pqts}\delta_{rt}+\frac{1}{2}v_{pqus}\delta_{ru}+\frac{1}{2}v_{pqrt}\delta_{st}+\frac{1}{2}v_{pqru}\delta_{su}\nonumber \\
+ & v_{turs}\delta_{pu}\delta_{qt}+v_{utrs}\delta_{pt}\delta_{qu}+v_{tqus}\delta_{pu}\delta_{rt}+v_{uqts}\delta_{pt}\delta_{ru}\nonumber \\
+ & v_{tqru}\delta_{pu}\delta_{st}+v_{uqrt}\delta_{pt}\delta_{su}+v_{ptus}\delta_{qu}\delta_{rt}+v_{puts}\delta_{qt}\delta_{ru}\nonumber \\
+ & v_{ptru}\delta_{qu}\delta_{st}+v_{purt}\delta_{qt}\delta_{su}+v_{pqtu}\delta_{ru}\delta_{st}+v_{pqut}\delta_{rt}\delta_{su}.\nonumber
\end{align}
These equations are sufficient to calculate the CC contribution to
the orbital gradient and diagonal Hessian in \eqref{E-DM}. The contributions
from the reference energy
\begin{align*}
E_{0}= & \sum_{i}h_{ii}+\frac{1}{2}\sum_{ij}\braket{ij||ij}\\
= & \sum_{i}f_{ii}-\frac{1}{2}\sum_{ij}\braket{ij||ij}
\end{align*}
are known to be\cite{Head-Gordon1988a}
\begin{align*}
\frac{\partial E_{0}}{\partial\Delta_{ai}}= & 2f_{ai}\\
\frac{\partial E_{0}}{\partial\Delta_{ai}^{2}}= & 2\left(f_{aa}-f_{ii}\right)+2\braket{ai||ia}.
\end{align*}
\bibliographystyle{tfo}
\bibliography{citations}

\end{document}